
\def\be{{\bf e}}
\def\half{{1/2}}
\def\refcpt{1}
\def\refKmes{2}
\def\refqgcpt{3}
\def\refHH83{4}
\def\refSmoot{5}
\def\refhaha85{6}
\def\refH85{7}
\def\refHW85{8}
\def\refPa85{9}
\def\refLa88{10}
\def\refLS87{11}
\def\refHalHar{12}
\def\refLy92{13}
\def\refHaL90{14}
\def\refli63{15}
\def\refge78{16}
\def\refG74{17}
\def\refLaL91{18}
\def\refgr89{19}
\def\refwa88{20}
\def\refbar{21}
\def\refmuk91{22}
\magnification 1200

\centerline{{\bf The Origin of Time Asymmetry}}
\vskip .5 true in
\centerline{$^{\dag}$S.W.Hawking\footnote{$^{{*}}$}{
Email:\ swh1@phx.cam.ac.uk.}, $^{\dag ,\ddag}$R.Laflamme\footnote{$^{{**}}$}{
Email:\ rl104@phx.cam.ac.uk.} \& $^{\dag}$G.W. Lyons\footnote{$^{{***}}$}{
Email:\ gwl10@phx.cam.ac.uk.}}
\vskip .5 true in

\centerline{$^{\dag}$Department of Applied Mathematics and
Theoretical Physics} \centerline{University of Cambridge}
\centerline{Silver Street, Cambridge }
\centerline{U.K., CB3 9EW}
\vskip .15 true in
\centerline{\&}
\vskip .15 true in
\centerline{$^{\ddag}$Theoretical Astrophysics, T-6, MSB288}
\centerline{Los Alamos National Laboratory}
\centerline{Los Alamos, NM 87545}
\centerline{USA}

\vskip 1 truein
\centerline{\bf Abstract}
\vskip .3 true in
\baselineskip 28truept

It is argued that the observed Thermodynamic Arrow of Time must arise from
the boundary conditions of the universe. We analyse the consequences of
the no boundary proposal, the only reasonably complete set of boundary
conditions that has been put forward. We study perturbations of a
Friedmann model containing a massive scalar field but our results should
be independent of the details of the matter content. We find that
gravitational wave perturbations have an amplitude that remains in the
linear regime at all times and is roughly time symmetric about the time of
maximum expansion. Thus gravitational wave perturbations do not give rise
to an Arrow of Time. However density perturbations behave very
differently. They are small at one end of the universe's history, but grow
larger and become non linear as the universe gets larger. Contrary to an
earlier claim, the density perturbations do not get small again at the
other end of the universe's history. They therefore give rise to a
Thermodynamic Arrow of Time that points in a constant direction while the
universe expands and contracts again. The Arrow of Time does not reverse
at the point of maximum expansion. One has to appeal to the Weak Anthropic
Principle to explain why we observe the Thermodynamic Arrow to agree with
the Cosmological Arrow, the direction of time in which the universe is
expanding.
\vfill\eject

\proclaim 1) Introduction.

The laws of physics do not distinguish the future from the past direction
of time. More precisely, the famous CPT theorem$^{\refcpt}$ says that the laws
are invariant under the combination of charge conjugation, space inversion and
time reversal. In fact effects that are not invariant under the combination
CP are very weak, so to a good approximation, the laws are invariant under
the time reverseal operation T alone. Despite this, there is a very obvious
difference between the future and past directions of time in the universe
we live in. One only has to see a film run backward to be aware of this.

There are are several expressions of this difference. One is the so-called
psychological arrow, our subjective sense of time, the fact that we
remember events in one direction of time but not the other. Another is the
electromagnetic arrow, the fact that the universe is described by  retarded
solutions of Maxwell's equations and not advanced ones. Both of these
arrows can be shown to be consequences of the thermodynamic arrow, which
says that entropy is increasing in one direction of time. It is a non
trivial feature of our universe that it should have a well defined
thermodynamic arrow which seems to point in the same direction everywhere
we can observe. Whether the direction of the thermodynamic arrow is also
constant in time is something we shall discuss shortly.

There have been a number of attempts to explain why the universe should
have a thermodynamic arrow of time at all. Why shouldn't the universe be
in a state of maximum entropy at all times? And why should the direction
of the thermodynamic arrow agree with that of the cosmological arrow, the
direction in which the universe is expanding? Would the thermodynamic
arrow reverse, if the universe reached a maximum radius and began to
contract?

Some authors have tried to account for the arrow of time on the basis of
dynamic laws. The discovery that CP invariance is violated in the decay of
the $K \sp o $ meson$^{\refKmes}$, inspired a number of such attempts but it is
now generally recognized that CP violation can explain why the universe
contains baryons rather than anti baryons, but it can not explain the
arrow of time. Other authors$^{\refqgcpt}$ have questioned whether quantum
gravity might not violate CPT, but no mechanism has been suggested. One would
not be satisfied with an ad hoc CPT violation that was put in by hand.

The lack of a dynamical explanation for the arrow of time suggests that it
arises from boundary conditions. The view has been expressed that the
boundary conditions for the universe are not a question for Science, but
for Metaphysics or Religion. However that objection does not apply if
there is a sense in which the universe has no boundary. We shall therefore
investigate the origin of the arrow of time in the context of the no
boundary proposal of Hartle \& Hawking$^{\refHH83}$. This was formulated in
terms of Einsteinian gravity which may be only a low energy effective theory
arising from some more fundamental theory such as superstrings. Presumably
it should be possible to express a no boundary condition in purely string
theory terms but we do not yet know how to do this. However the recent
COBE observations$^{\refSmoot}$ indicate that the perturbations that lead to
the
arrow of time arise at a time during inflation when the energy density is about
$10^{-12} $ of the Planck density. In this regime, Einstein gravity should
be a good approximation.

In most currently accepted models of the early universe there is some
scalar field $\phi $ whose potential energy causes the universe to expand
in an exponential manner for a time. At the end of this inflationary
period, the scalar field starts to oscillate and its energy is supposed to
heat the universe and to be transformed into thermal quanta of other
fields. However this thermalisation process involves an implicit
assumption of the thermodynamic arrow of time. In order to avoid this we
shall consider a universe  in which the only matter field is a massive
scalar field. This will not be a completely realistic model of the
universe we live in because it will be effectively pressure free after the
inflationary period rather than radiation dominated. However it has the
great advantage of being a well defined model without hidden assumptions
about the arrow of time. One would expect that the existence and direction
of the arrow of time should not depend on the precise matter content of
the universe. We shall therefore consider a model in which the
 action is given by the Einstein-Hilbert action
$$
I_g = {1 \over 16\pi G} \int_{\cal M} d^4x \, {(-g)}^{\half} \, R
\ \ + \ \
{1 \over 8\pi G} \int_{\partial \cal M} d^3x \, {(h)}^{\half} \, K
\eqno(1.1)
$$
plus the massive scalar field action
$$
I_\Phi =
- \half \int_{\cal M}
d^4x \, {(-g)}^{\half} \, (  g^{\mu\nu}\partial_\mu \phi \partial_\nu \phi
                           + m^2 \phi^2 ).
\eqno(1.2)
$$

In accordance with the no boundary proposal, we shall take the quantum
state of the universe to be defined by a path integral over all compact
metrics with this action. This means that the wave function $ \Psi
[h_{ij}, \phi _0]$ for finding a three metric $h_{ij}$ and scalar field
$\phi _0$ on a spacelike surface $S$ is given by
$$
\Psi(h_{ij}, \phi_0) = \int_{\cal C} d[g_{\mu\nu}]d[\phi] \,
                  {\rm e}^{-I_e[g_{\mu\nu},\phi]}
\eqno (1.3)
$$
where the path integral is taken over all metrics and scalar fields on compact
manifolds $M$ with boundary $S$ that induce the given values on the boundary.
In
general the metrics in the path integral will be complex rather than
purely Lorentzian or purely Euclidean.

There are a number of problems in defining a path integral over all
metrics, two of which are:

 \item{(1)} The path integral is not perturbatively renormalisable.

 \item{(2)} The Einstein Hilbert action is not bounded below.

These difficulties may indicate that Einstein gravity is only an effective
theory. Nevertheless, for the reasons given above we feel the saddle point
approximation to the path integral should give reasonable results. We shall
therefore endeavour to evaluate the path integral at stationary points of
the action, that is at solutions of the Einstein equations. These
solutions will be complex in general.

The behaviour of perturbations of a Friedmann model according to the no
boundary proposal was first investigated by Halliwell \&
Hawking$^{\refhaha85}$ and we shall adopt their notation. The perturbations
are  expanded in hyperspherical harmonics. There are three kinds of
harmonics.

\item {(1)} Two degrees of freedom in tensor harmonics. These are gauge
invariant and correspond to gravitational waves.

\item {(2)} Two degrees of freedom in vector harmonics. In the model in
question they are pure gauge.

\item {(3)} Three degrees of freedom in scalar harmonics. Two of them
correspond to gauge degrees of freedom and one to a physical density
perturbation.

One can estimate the wave functions for the perturbation modes by
considering complex metrics and scalar fields that are solutions of the
Einstein equations whose only boundary is the surface $S$. When $S$ is a
small three sphere, the complex metric can be close to that of part of a
Euclidean four sphere. In this case the wave functions for the tensor and
scalar modes correspond to them being in their ground state. As the three
sphere $S$ becomes larger, these  complex metrics change continuously to
become almost Lorentzian. They represent universes with an initial period
of inflation driven by the potential energy of the scalar field. During
the inflationary phase the perturbation modes remain in their ground
states until their wave lengths become longer than the horizon size. The
wave function of the perturbations then remains frozen until the horizon
size increases to be more than the wave length again during the matter
dominated era of expansion that follows the inflation. After the wave
lengths of the perturbations come back within the horizon, they can be
treated classically.

This behaviour of the perturbations can explain the existence and direction
of the thermodynamic arrow of time. The density perturbations when they
come within the horizon are not in a general state but in a very special
state with a small amplitude that is determined by the parameters of the
inflationary model, in this case, the mass of the scalar field. The recent
observations by COBE indicate this amplitude is about $10^{-5}$. After the
density perturbations come within the horizon, they will grow until they
cause some regions to collapse as proto-galaxies and clusters. The
dynamics will become highly non linear and chaotic and the coarse grained
entropy will increase. There will be a well defined thermodynamic arrow of
time that points in the same direction everywhere in the universe and
agrees with the direction of time in which the universe is expanding, at
least during this phase.

The question then arises: If and when the universe reaches and maximum
size, will the thermodynamic arrow reverse? Will entropy decrease and the
universe become smoother and more homogeneous during the contracting
phase? In reference [\refH85] it was claimed that the no boundary proposal
implied that the thermodynamic arrow would reverse during the contraction.
This is now recognized to be incorrect but it is instructive to consider
the arguments that led to the mistake and see why they do not apply. The
anatomy of error is not ruled by logic but there were three arguments
which together seemed to point to reversal:

\item {(1)} The no boundary proposal implied that the wave function of the
universe was invariant under CPT.

\item {(2)} The analogy between spacetime and the surface of the Earth
suggested that if the North Pole were regarded as the beginning of the
universe, the South Pole should be its end. One would expect conditions to
be similar near the North and South Poles. Thus if the amplitude of
perturbations was small at early times in the expansion, it should also be
small at late times in the contraction. The universe would have to get
smoother and more homogeneous as it contracted.

\item {(3)} In studies of the Wheeler Dewitt equation on minisuperspace
models$^{\refHW85}$ it was thought that the no boundary condition implied that
$ \Psi (a) \rightarrow 1$ as the radius $ a \rightarrow 0$. In the case of a
Friedmann model with a massive scalar field, this seemed to imply that the
classical solutions that corresponded to the wave function through the WKB
approximation would bounce and be quasi-periodic. This could be true only
if the solutions were restricted to those in which the perturbations
became small again as the universe contracted.

Page$^{\refPa85}$ pointed out that the first argument about  the CPT invariance
of the wave function  didn't imply that the individual histories had to be CPT
symmetric, just that if the quantum state contained a particular history,
then it must also contain the CPT image of that history with the same
probability. Thus this argument didn't necessarily imply that the
thermodynamic arrow reversed in the contracting phase. It would be equally
consistent with CPT invariance for there to be histories in which the
thermodynamic arrow to pointed forward during both the expansion and
contraction, and for there to be other histories with equal probability in
which the arrow was backward. With a relabelling of time and space
directions and of particles and antiparticles, these two classes of
histories would be physically identical. Both would correspond to a steady
increase in entropy from one end of time, which can be labelled the Big
Bang, to the other end, which can be labelled the Big Crunch.

The second argument, about the north and south poles being similar, is
really a confusion between real and imaginary time. It is true that there
is no distinction between the positive and negative directions of time. In
the Euclidean regime, the imaginary time direction is on the same footing
as spatial dimensions. So one can reverse the direction of imaginary time
by a rotation. Indeed, this is the basis of the proof that the no boundary
quantum state is CPT invariant. But as noted above, this does not imply
that the individual histories are symmetric in real time or that the Big
Crunch need be similar to the Big Bang.

The third argument, that the boundary condition for the Wheeler Dewitt
equation should be $\Psi \rightarrow 1$ for small three spheres $S$ in a
homogeneous isotropic mini superspace model, was the one that really led
to the error of suggesting that the arrow of time reversed. The motivation
behind the adoption of this boundary condition was the idea that the
dominant saddle point in the path integral for a very small three sphere
would be a small part of a Euclidean four sphere. The action for this
would be small. Thus the wave function would be about one irrespective of
the value of the value of the scalar field. With this boundary condition,
the mini superspace Wheeler Dewitt equation gave a wave function that was
constant or exponential for small radius, and which oscillated rapidly for
larger radius. From the WKB approximation one could interpret the
oscillations as corresponding to Lorentzian geometries. That fact that the
oscillating region didn't extend to very small radius was taken to
indicate that these Lorentzian geometries wouldn't collapse to zero radius
but would bounce. Thus they would correspond to quasi-periodic oscillating
universes. In such universes, the perturbations would have to obey a
quasi-periodic boundary condition and be small whenever the radius of the
universe was small. Otherwise the universe would not bounce. This would
mean that the thermodynamic arrow would have to would reverse during the
contraction phase so that the perturbations were small again at the next
bounce.

This boundary condition on the wave function became suspect when
Laflamme$^{\refLa88 ,\refLS87}$ found other minisuperspace models in which a
bounce was not possible. Then Page$^{\refPa85}$ pointed out that for small
three
surfaces $S$, there was another saddle point that could make a significant
contribution to the wave function. This was a complex metric that started
almost
like half of a Euclidean four sphere and was followed by an almost Lorentzian
metric that expanded to a maximum radius, and then collapsed to the small three
surface $S$. The long Lorentzian period would give the action of these
metrics a large imaginary part. This would lead to a contribution to the
wave function that oscillated very rapidly as a function of the radius of
the three surface $S$ and the value of the scalar field on it. Thus the
boundary condition of the Wheeler Dewitt equation wouldn't be exactly
$\Psi \rightarrow 1$ as the radius tends to zero. There would also be a
rapidly oscillating component of the wave function.

As before, the wave functions for perturbations about the Euclidean saddle
point metric would be in their ground states. But there is no reason for
this to be true for perturbations about the saddle point metric with a
long Lorentzian period that expanded to a large radius and then contracted
again.

To find out what the wave functions for perturbations in the contracting
phase are, one has to solve the relevant Schroedinger equation during the
expansion and contraction. This we do in sections (3.1) and (3.2). We find that
the tensor modes have wave functions that correspond to gravitational
waves that oscillate with an adiabatically varying amplitude. This
amplitude will depend on the radius of universe. It will be the same at
the same radius in the expanding and contracting phases and it will be
small compared to one whenever the wave length is less than the horizon
size. Thus these gravitational wave modes will not become non linear and
will not give rise to a thermo dynamic arrow of time.

By contrast, scalar modes between the Compton wave length of the scalar
field and the horizon size won't oscillate but will have power law
behaviour. There are two independent solutions of the perturbation
equations, one which grows and one which decreases with time. The boundary
condition provided by the no boundary proposal picks out the solution that
is a small perturbation about the Euclidean saddle point for small three
spheres. It does not require that the perturbation about the saddle point
with a long Lorentzian period remains small. So the no boundary proposal
picks out the solution of the density perturbation equation that starts
small but grows during the expansion and continues to grow during the
contraction. At some point during the expansion, the amplitude will grow
so large that the linearized treatment will break down. This however does
not prevent one using linear perturbation theory  to draw conclusions
about the thermodynamic arrow of time. The arrow of time  is determined by
when the evolution becomes non linear. The linear treatment and the no
boundary proposal enable one to say that this will happen during the
expansion. After that the evolution will become chaotic and the coarse
grained entropy will increase. It will continue to increase in the
contracting phase because there is no requirement that the perturbations
become small again as the universe shrinks. Thus the thermodynamic arrow
will not reverse. It will point the same way while the universe expands
and contracts.

The thermodynamic arrow will agree with the cosmological arrow for half the
history of the universe, but not for the other half. So why is it that we
observe them to agree? Why is it that entropy increases in the direction
that the universe is expanding? This is really a situation in which one can
legitimately invoke the weak anthropic principle because it is a question
of where in the history of the universe conditions are suitable for
intelligent life. The inflation in the early universe implies that the
universe will expand for a very long time before it contracts again. In
fact, it is so long that the stars will have all burnt out and the baryons
will have all decayed. All that will be left in the contracting phase will
be a mixture of electrons, positrons, neutrinos and gravitons. This is not
a suitable basis for intelligent life.

The conclusion of this paper is that the no boundary proposal can explain
the existence of a well defined thermodynamic arrow of time. This arrow
always points in the same direction. The reason we observe it to point in
the same direction as the cosmological arrow is that conditions are
suitable for intelligent life only at the low entropy end of the
universe's history.

\proclaim 2) The Homogeneous Model.

In this section we review the homogeneous model with metric
$$
ds^{2}=\sigma ^{2}(-N(t)^{2}dt^{2}+a(t)^{2}d\Omega ^{2}_{3})
\eqno(2.1)
$$
where $\sigma ^{2} = 2/(3\pi m^{2}_{p})$, $N$ is the lapse function, $a$ is the
scale factor and $d\Omega ^{2}_{3}$ is
the standard 3-sphere metric. Expressing the scalar field as $\sqrt {2}
\pi \sigma \phi $ with the quadratic potential $2\pi ^{2} \sigma ^{2} m^{2}
\phi ^{2}$, the Lorentzian action is
$$
I=-{1\over2}\int dt Na^{3}[{{\dot a}^{2} \over N^{2}a^{2}}
-{1 \over a^{2}}-{{\dot \phi }^{2} \over N^{2}}
+m^{2}\phi ^{2}]
\eqno(2.2)
$$
where the dot denotes derivative with respect to Lorentzian FRW time (if not
explicitely stated  throughout the paper time derivative are Lorentzian). There
are no time derivatives of the lapse function $N$ in this action; it is a
lagrange multiplier.  Varying the action with respect to $N$ leads to the
constraint  $$ H={N\over 2a^{3}}[-a^{2} \pi ^{2}_{a} +\pi^{2}_{\phi } -
  a^{4}(1-a^{2}m^{2}\phi ^{2})] =0
\eqno(2.3)
$$
where the momenta $\pi_{a}$ and $\pi_{\phi }$ are defined as
$$
\pi _{a}=-{a \over N}\dot a \qquad
{\rm and}
\qquad \pi _{\phi }={a^{3} \over N}\dot \phi
\eqno(2.4)
$$
and $H$ is the Hamiltonian. This constraint is a consequence of the
invariance under time reparametrization.
Varying the action with respect to the field $\phi$ we obtain the reduced
Klein-Gordon equation
$$
N{d \over dt}({{\dot \phi } \over N }) +3{\dot a \over a}{\dot \phi} +
N^{2}m^{2}\phi ^{2}=0,
\eqno(2.5)
$$
This latter equation together with the Hamiltonian constraint $H=0$, is
sufficient to describe the classical dynamics.  The  second order equation  for
$a$ can be derived from these equations. In the inhomogeneous model there are
also momentum constraints, but these are trivially satisfied in the
homogeneous background.

The quantum theory is obtained by replacing the different variables by
operators.  We will follow the Dirac method and impose the classical
constraints
as quantum operators.  The Hamiltonian constraint thus becomes
$$
[a^{2}{\partial ^{2} \over \partial a^{2}}-{\partial ^{2} \over \partial
\phi ^{2}} -a^{4}(1-a^{2}m^{2}\phi ^{2})]\Psi_0 (a,\phi ) = 0
\eqno(2.6)
$$
and is called the Wheeler-DeWitt equation.  The solution of this equation
$\Psi _0(a,\phi)$ is the wave function of the universe. There is a  factor
ordering  ambiguity, but it is not important for the conclusions of our paper
which rely on the classical limit.

In this paper we investigate the predictions of the no-boundary proposal
in a model where small inhomogeneities are taken into account.
In order to impose this proposal we return to a path integral formulation of
the
wavefunction.  It is very hard to calculate this path integral exactly.
However we can have a good idea of the resulting wave function by using a
saddlepoint approximation
$$
\Psi(h_{ij}, \phi) \approx C\
                  {\rm e}^{-I^{sp}_{_E}[g_{\mu\nu}\Phi]}
\eqno (2.7)
$$
where $C$ is a prefactor and $I^{sp}_{_E}$ is the Euclidean saddle-point
action.
In this approximation it is clear how to impose the proposal of Hartle
and Hawking.  The regularity condition is imposed on the (complex)
saddlepoints of the path integral. The semiclassical approximation to the path
integral can then be used to estimate the wavefunction.

One of the problems in using the semiclassical approximation in this model is
that we cannot simply deform the complex metric into purely real Euclidean and
real Lorentzian sections, for real arguments of the wavefunction. This could
only
be achieved in this model if the time derivatives of both $a$ and $\phi $
vanish
simultaneously on the Euclidean axis,$^{\refHalHar}$ which is not possible as
$\phi $ increases monotonically if the no boundary condition is imposed.
Therefore we must solve the background equations of motion for complex values
of
time and physical variables, obtaining complex solutions which satisfy the no
boundary proposal and have the given $a$ and $\phi $ on the final hypersurface.
The no boundary proposal imposes the boundary conditions at one end of the four
geometry $$
a=0 \qquad {da \over d\tau }=1 \qquad
{d\phi \over d\tau }=0 \qquad \phi = \phi _{0}
\eqno(2.8)
$$
thus we only have the freedom to choose the (complex) value of $\phi $ at the
origin of complex time $\tau$.

Lyons$^{\refLy92}$ found that there were many contours in the complex
time  plane which induced real endpoints $a$ and $\phi$.
Some possibilities are obtained by choosing the initial value of $\phi $ to
have an imaginary part much smaller than the real part such that
$\phi^{Im}_{0} \approx -(1+2n)\pi /6\phi _{0}^{Re}$ (for integer $n$).  In
this paper we will only investigate the case $n=0$.

For small $a$ the complex metric can effectively be considered as a small real
Euclidean section, with $\phi_0$ approximately real, described by
$$
\phi \approx \phi _{0} \qquad {\rm and} \qquad a \approx {1\over m\phi
_{0}}\sin m\phi _{0}\tau
\eqno(2.9)
$$
where $\tau $ is the Euclidean time. When we consider gravitons below, it is a
good approximation to assume the following behaviour for the radius $a$ when
$\phi_0>1$.  For small $a$ ($ <m\phi_0$) the background is part of an
Euclidean 4-sphere
$$
a \approx {1\over m\phi_0 \cosh \eta_{_E}}  \qquad -\infty<\eta_{_E}<0.
\eqno (2.10)
$$
The Euclidean conformal time is given by $\eta _{E} = \int d\tau /a $. Although
$\eta _E $ has semi-infinite  range notice that the proper
distance is finite.   The radius $a$ starts at zero and increases to a maximum
value of $1/m\phi_0$, the equator of the 4-sphere. For larger $a$, the
saddle point is well approximated by de Sitter space
$$
a \approx {1\over m\phi_0 \cos \eta}  \qquad 0<\eta< {\pi \over
2}-\delta_e \eqno (2.11)
$$
where $\eta$ is the analytic continuation of $\eta_{_E}=i\eta$.  The universe
is
then in an inflationary era. In terms of comoving time:
$$
\phi \approx \phi _{0}-{mt \over 3} \qquad
{\rm and}
\qquad a \approx {1 \over m\phi
_{0}}e^{m\phi_{0}t-{1\over6}m^{2}t^{2}}    \eqno(2.12)
$$
where $t$ is the analytic continuation of $\tau$ in the Lorentzian region.

The action is given by
$$
I_e \approx  -{1\over 3m^2\phi^2_0}\Big ( 1 -  (1-m^2\phi^2_0a^2)^{3/2} \Big).
\eqno (2.13)
$$
For large $a$ ($\gg 1/m\phi_0$),  the saddle point  will have a large imaginary
part.  The wave function will therefore be of WKB type.
After a suitable coarse graining,$^{\refHaL90}$  we can
associate the phase of the wave function to the Hamilton-Jacobi function of
general relativity.  When this is possible we will assume that the universe
behaves essentially classically. The wave function will be associated to
the family of classical Lorentzian trajectories described by the
Hamilton-Jacobi
function.

Meanwhile the scalar field is decreasing and
inflation will end at $\eta=\pi /2 -\delta _e $ when the scalar field reaches a
value around unity, at which point the value of $a$ will be
$a_{e} \approx (1/m\phi _{0})\exp (3\phi _{0}^{2}/2)$. $\delta_e$ is given by
the implicit relation $\delta_e\approx \exp({-3(\phi_0)^2/2})$.  For
$\phi_0 > 1$, we have $\delta_e \ll 1$.
When $\eta>\pi /2 - \delta_e$, the scalar field oscillates and behaves
essentially as a pressureless fluid (i.e. dust):
$$
\phi \approx {1 \over m}({a_{\rm max} \over a^{3}})^{1/2} \cos(mt).
\eqno(2.14)
$$

 The scale factor of the universe is then well described by
$$
a \approx a_m \sin^2({\pi /2 - 3\delta_e - \eta \over 2})  \qquad {\pi \over 2}
-\delta_e <\eta \eqno (2.15)
$$
where the constants have been chosen to ensure a smooth transition between the
inflationary  and dust era. The universe will therefore expand to a maximum
radius  $a_m\approx m^2 a_e^3
\approx \exp({9(\phi_0)^2/2})/m({\phi_0})^{3}$ and recollapse.
It will be convenient later on to redefine the origin of conformal time during
the dust-like era by setting $\eta_d = \eta -\pi /2 + 3\delta_e$.
 The scale factor will then evolve as
$$
a \approx a_{\rm max}\sin ^2 {\eta_d \over 2} \qquad 0 < \eta_d < 2\pi
\eqno(2.16)
$$
 Figure 1 depicts a typical classical trajectory
corresponding to the no-boundary proposal.

\proclaim 3) Inhomogeneous Perturbations.

Let us now consider the behaviour of small perturbations around the
the homogeneous model described in the previous section.
We write the metric as
$$
g_{\mu \nu }(t,{\bf x}) =  g_{\mu \nu }(t) + \delta g_{\mu \nu }(t,{\bf x}).
\eqno (3.1)
$$
The background part $ g_{\mu \nu }(t)$ was decribed in the previous section by
the line element (2.1).

One can decompose a general perturbation $\delta g_{\mu \nu }$ of a
Robertson-Walker background metric into scalar ($Q^{n}_{lm}$), vector
($(P_{i})^{n}_{lm},(S_{i}^{o,e})^{n}_{lm}$) and tensor
($(P_{ij})^{n}_{lm},(S_{ij}^{o,e})^{n}_{lm},(G_{ij}^{o,e})^{n}_{lm}$)
harmonics.  This classification originates from the way
they transform under rotations of the 3-sphere.
These harmonics are constructed from the scalar, vector and tensor
eigenfunctions
of the Laplacian on the 3-sphere, viz. $Q^{n}_{lm}$, $(S^{o,e}_{i})^{n}_{lm}$
and $(G^{o,e}_{ij})^{n}_{lm}$. More details and properties of these harmonics
are given in refs. [\refli63,\refge78].

We can expand the inhomogeneous perturbations of the metric in terms of these
harmonics  (where the index $n$ should be thought of as a shorthand for
$nlm$ and ${o,e}$).
The tensor perturbations are:
$$
\delta g_{\mu \nu }^{(t)} = \sum _{n} a^{2}\left(\matrix{0&0\cr
0&2d_{n}G^{n}_{ij}\cr} \right) .
\eqno(3.2)
$$
$G^{n}_{ij}$ are the transverse traceless tensor harmonics.
The vector perturbations are:
$$
\delta g_{\mu \nu }^{(v)} = \sum _{n}{a^{2}\over \sqrt{2}}
\left(\matrix{0&j_{n}S^{n}_{i}\cr
       j_{n}S^{n}_{i}&2c_{n}S^{n}_{ij}\cr}
\right)
\eqno(3.3)
$$
where the $S^{n}_{ij}=S^{n}_{i|j}+S^{n}_{j|i}$ are obtained from the
transverse vector harmonics $S^{n}_{i}$.
The scalar perturbations of the
$$
\delta g_{\mu \nu }^{(s)} = \sum _{n}{a^{2} \over \sqrt{6}}
\left(\matrix{-2N_{0}^{2}g_{n}Q^{n}&k_{n}P^{n}_{i}\cr
               k_{n}P^{n}_{i}&2a_{n}\Omega
                _{ij}Q^{n}+6b_{n}P^{n}_{ij}\cr}\right)
\eqno(3.4)
$$
where the $P^{n}_{i}=Q^{n}_{|i}/(n^{2}-1)$ and $P^{n}_{ij}=\Omega
_{ij}Q^{n}/3 + Q^{n}_{|ij}/(n^{2}-1)$ are obtained from the scalar
harmonics $Q^{n}$.
We must also take into account the scalar perturbations of the scalar field:
$$
\delta \phi = \sum _{n} {1\over \sqrt{6}}f_{n}Q^{n}  .
\eqno(3.5)
$$

This expansion is in effect a Fourier transform adapted to the symmetry of the
FRW  background.  The coefficients $a_n, b_n, c_n, d_n, f_n, g_n, j_n$ and $k_n
$are functions of time, but not of the spatial  coordinates of the three-sphere
hypersurfaces. Spatial information is encoded  in the harmonics.

In [\refhaha85] the action (1.3) and (1.4) was expanded to second order
around the homogeneous model.  In appendix A,
we have reproduced it with the equations of motion for the various Fourier
coefficients.  After examining the perturbed Lagrangians (A.2) and (A.3) we
find
that the different types of harmonics decouple from each other. Their wave
functions will therefore separate so we can write
$$
\Psi_n(a,\phi,a_n,b_n,c_n,d_n,f_n)=\psi^s_n(a,\phi ,a_n,b_n,f_n)
\psi^v_n(a,\phi,c_n)\psi^t_n(a,\phi,d_n)
\eqno (3.6)
$$
It is thus possible to investigate them separately. We will study the
tensor and scalar modes in the next two subsections.  For the vector modes
there
are only two variables $c_{n}$ and $j_{n}$.  The latter one however is a
Lagrange multiplier and thus induces a constraint for the
only variable left. Thus we find that the vector degrees of freedom are pure
gauge and will only  contribute to the phase of the total wave function.

\proclaim 3.1) Linear Gravitons.

Linear gravitons are the transverse and traceless part of the 3-metric and
are described by the variables $d_n$ in the above notation.  Using the
background
equation of motion we can derive the equation$^{\refG74}$
$$
d_{n}''+2{\cal H}d_{n}' +(n^{2}-1)d_{n} = 0.
\eqno(3.7)
$$
for the modes $d_n$. Here the derivatives are with respect to Lorentzian
conformal time and ${\cal H} = a'/a$.
The gravitons are decoupled from the scalar and vector-derived  tensor
harmonics
and depend only on the behaviour of the background.

We will calculate the wave function for the graviton modes using a
saddle-point approximation, assuming the background wave function (2.7)
and saddle-point action (2.13).
The tensor part of the wave function (see 3.6) can be written as
$$
\eqalign{
\psi^t_n(a, \phi_0,d_n) &=  \int [d d_n] {\bf e}^{-(I_E)}\cr
       &\approx C \be^{-(I^{ext}_E)}\cr}
\eqno (3.8)
$$
where $C= (\delta^2 (I^{ext}_E)/\delta d^i_n \delta d^f_n)^{1/2}$
is the prefactor assuming the flat spacetime measure.

The Euclidean action for a mode $d_n$ calculated along an extremising path is
given by the boundary term
$$
I^{ext}_E = ({a^2 d_n d^\prime_n \over 2} + 2aa^\prime d^2_n )
                                           \Big |^{\eta^f_{_E}}_{\eta^i_{_E}}
\eqno (3.9)
$$
where $\eta_{_E}$ is the Euclidean time, a function of the background variables
$a$ and  $\varphi_0$ as described in [\refLaL91].  It is possible to
rewrite this action in terms of values of the field on the boundary
$d^i_n,d^f_n$ and solutions of the classical equation $p_n$
$$
{d\over d\eta_{_E}}a^2 {d\over d\eta_{_E}} p_n - (n^2-1) a^2p_n =0
\eqno (3.10)
$$
evaluated on the boundary.
The regularity condition for the no-boundary proposal implies that $d_n$ must
vanish when the 3-geometry shrinks to zero and this implies that the action
will
have the form
$$
I^{ext}_E = A d_n^2 = {a^2\over 2}
      ({p_n^\prime\over p_n}+4{a^\prime\over a}) d^2_n.
\eqno (3.11)
$$
In regions of configuration space where the universe is
Lorentzian, the appropriate analytic continuation of (3.11)
should be taken.

It is possible to find a good analytical approximation for $p_n$ and thus
of the wave function  using (3.8) and (3.11) and assuming that the background
is described by equations (2.10), (2.11) and (2.15).  The $p_n$ are
approximately
$$
\eqalign{
p_n &\propto (\cosh\eta_{_E}-{\sinh\eta_{_E}\over n} ) \be^{n\eta_{_E}},
\ \  -\infty<\eta_{_E} <0,
              \ \ {\rm in\ the\ Euclidean\ region;}\cr
&\propto (\cos\eta+i{\sin\eta\over n}) \be^{-in\eta},
\ \ \ \ \ \ \ \ \ \ \ 0<\eta\ < {\pi \over 2}-\delta_e,
              \ \ {\rm in\ the\ inflationary\ era;}\cr
&\propto \Big({\cos[n(\eta-3\pi /2 +3\delta_e)]\over\cos^2[(\eta-3\pi /2
+3\delta_e)/2]}
      -{\sin[(\eta-3\pi /2 +3\delta_e)/2] \sin[n(\eta-3\pi /2 +3\delta_e)]
        \over 2n \cos^3[(\eta-3\pi /2 +3\delta_e)/2]}\Big)\cr
& \ \ \ \ \ \ \ \ \ \ \ \ \ \ \ \ \ \ \ \ \ \ \ \ \ \ \ \ \ \ \ \ \ \ \ \ \ \ \
  \ \  \
  {\pi\over 2}-\delta_e <\eta\ ,
   \ \ {\rm in\ the\ dust-like\ phase.}\cr}
\eqno (3.12)
$$
Modes with $n\delta_e \ll 1$ are those with wavelengths much
larger than the Hubble radius at the  end of inflation.  At the onset of
inflation they are in their ground state and thus oscillate adiabatically.
These modes will no longer oscillate adiabatically when they leave the
Hubble radius during inflation. However all modes will re-enter the Hubble
radius during the dust era when $n \approx \tan[(\eta -\pi /2 +3\delta_e)/2]$
and
start oscillating adiabatically again. Modes with $n\delta_e \gg 1$ oscillate
adiabatically throughout the evolution. All the modes oscillate around the time
of maximum expansion, and even if some do not  have a phase which is exactly
time
symmetric, their amplitudes are.

The variance squared of the field and its momenta for modes with
$n\delta_e\ll 1$   around the time of maximum expansion  are given by
$$
\langle d^2_n \rangle = {1\over 2(A^*+A)}
    \approx {(1+2\gamma\cos(2n\eta) + \gamma^2)\over 2 na^2(1-\gamma^2)}
\eqno (3.13)
$$
$$
\langle \pi^2_{d_n} \rangle = {A^* A\over 2(A^*+A)}
    \approx {na^2\over 2}{(1-\gamma^2)^2+4\gamma^2\sin^2(2n\eta)\over
     (1+2\gamma\cos(2n\eta) +\gamma^2)(1-\gamma^2)}
\eqno (3.14)
$$
and
$$
\langle d_n\pi_{d_n}+\pi_{d_n}d_n \rangle={i(A-A^*)\over(A+A^*)}
    \approx {4\gamma\sin(2n\eta)\over (1-\gamma^2)}
\eqno (3.15)
$$
where $\gamma=1-n^2\delta^2_e/2$.
The expectation value of the Hamiltonian
$$
H_n={1\over 2a^3}\Big[ \pi^2_{d_n}
   + 4(d_n\pi_{d_n}+\pi_{d_n}d_n)a\pi_a
   + d^2_n [10a^2\pi^2_a +6\pi^2_{\phi}- 6a^6m^2\phi^2 + (n^2+1)a^4]
               \Big]
\eqno (3.16)
$$
is
$$
\eqalign{
\langle H_n \rangle &\approx {n\over a}
                             \ \ {\rm at\ the\ onset\ of \ inflation}\cr
&\approx {n\over a n^2\delta^2_e}\ \ {\rm near\ the\ maximum\ expansion}.\cr}
\eqno (3.17)
$$
This shows that modes start in their ground state before the onset of inflation
and get excited during inflation and the dust phase.

A useful way to gain information about this state is to investigate the Wigner
function
$$
{\cal F}(\bar d_n, \bar\pi_n) = {1\over 2\pi}
\int d \Delta \be^{-2i \bar\pi \Delta}
          \psi^*(\bar d_n-\Delta) \psi(\bar d_n +\Delta).
\eqno (3.18)
$$
The  Wigner function gives an idea of the phase space probability distribution
of possible classical perturbations (once decoherence has occured).  For the
wave function (3.8)  with action (3.11), it is given by
$$
{\cal F}(\bar d_n, \bar\pi_n) = {A+A^*\over 2\pi}
\exp{-( {4AA^*\over A+A^*}\bar d^2_n + {1\over A+A^*}\bar\pi^2_n
  -2i{A-A^*\over A+A^*}\bar d_n \bar\pi_n)}.
\eqno (3.19)
$$
At the onset  of inflation the Wigner  function is
a round Gaussian (factoring out the mode number and
the radius of the universe).   A mode with $n<\tan (\pi /2 -\delta_e)$ will
go outside the Hubble radius and have frozen amplitude and the Wigner function
will  then
become an ellipse elongated in the momentum direction.  When the mode comes
back
within the  Hubble radius it starts rotating with period $2\pi/n$ in phase
space.
This behaviour lasts until  $n\approx \tan\eta$ in the recontracting phase.
The
parameter characterizing the eccentricity of this ellipse is
called the squeezing and has been studied by Grishchuk \& Sidorov$^{\refgr89}$.

Typical classical perturbations $d^{cl}_n$ resulting from the above Wigner
function  are small at the onset of inflation.  Their amplitudes get frozen
when
they leave the Hubble radius. During this stage their energies increase.
The perturbations will start oscillating again with amplitude proportional to
$a^{-1}$ when they come back within the Hubble
radius in the dust phase.  They behave like
$$
d^{cl}_n \approx {\sin(n\eta +\epsilon)\over a n^{3/2}\delta_e}
\eqno (3.20)
$$
where $\epsilon$ is an unimportant phase depending on the details of the
matching of the $p_n$ functions in (3.12).
Around the time of maximum expansion the amplitude
of the graviton modes is symmetric and thus their arrow of time agrees with
the cosmological one.  Figure 2 depicts a typical classical
evolution of a linear graviton.

\vskip 1cm

\proclaim 3.2) Linear Scalar Perturbations.
\vskip 1truecm

{\bf \sl (a) Quantum Mechanics of the Physical Degree of Freedom}
\vskip 0.5truecm

We have seen that gravitons are adiabatic near the time of maximum
expansion so that their amplitude is time symmetric with respect to
that point.  This is not special to gravitons as the electromagnetic field,
massless or conformally coupled scalar fields will also be adiabatic.
In this  section we will show however that  perturbations of massive scalar
field
will not behave adiabatically at the time of maximum
expansion.

{}From the expansion (3.4) and (3.5) we see that there are five scalar degrees
of
freedom described by the time-dependent coefficients $a_n, b_n, f_n,k_n$ and
$g_n$.  However  the latter two appear as Lagrange multipliers in the
Lagrangians (A.2), (A.3) and induce two constraints so overall there is only
one
true scalar degree of freedom.  Without the presence of the scalar field the
scalar degrees of freedom would also be pure gauge. Care should be taken in the
treatment of the scalar perturbations in order to avoid gauge dependent
results. Let us first find the real degree of freedom.

Variations of the action with respect to the Lagrange multipliers $N$, $g_{n}$
and $k_{n}$ result in the Hamiltonian, linear Hamiltonian and momentum
constraints. In Dirac quantization, which we follow here, these constraints are
imposed as constraints on the quantum state.  The wave function therefore
depends only on a linear combination of the coffecients $a_n,b_n$ and $f_n$.
The momentum constraints ensures that the wavefunction is invariant under
diffeomorphisms of the spatial three-surfaces.  The Hamiltonian and linear
Hamiltonian constraints ensure time  reparametrization invariance of the wave
function.

Shirai and Wada$^{\refwa88}$ give an explicit form for the
wave function which automatically satisifies the momentum constraints.  These
are solved  by making the judicious change of variables
$$
\eqalign{
\tilde \alpha &= \alpha + {1\over 2}\sum_n a^2_n
                     - 2\sum_n{(n^2-4)\over (n^2-1)} b^2_n \cr
\tilde \phi &= \phi -3\sum_n b_n f_n .\cr}
\eqno (3.21)
$$
where $\alpha = \ln a$. Once this transformation has been performed, the
momentum constraints imply that the wave function is independent of the linear
combination $a_n-b_n$. In terms of the two degrees of freedom left, the
linear Hamiltonian constraint becomes
$$
\pi _{\phi }\pi _{f_n}-\pi _{\alpha }\pi _{s_n} + e^{6\alpha }m^2 \phi f_n +
K_n s_n =0  \eqno(3.22)
$$
where $s_n = a_n+b_n$ and $K_n = {1\over3}[(n^2-4)\pi _{\alpha }^2-(n^2+5)\pi
_{\phi }^2 - (n^2-4)e^{6\alpha }m^2 \phi ^2]$. The remaining gauge degree of
freedom can be eliminated by solving the linear Hamiltonian constraint using
the
canonical transformation
$$
\eqalign{\left(\matrix{y_n \cr
              z_n \cr}\right) &=
 \left(\matrix{K_n & e^{6 \alpha }m^2 \phi \cr
               \pi _{\phi } & \pi _{\alpha }}\right)
  \left(\matrix{s_n \cr
                f_n \cr}\right) \cr
\noalign {\bigskip }
\left(\matrix{\pi _{s_n} \cr
              \pi _{f_n} \cr}\right) &=
 \left(\matrix{K_n & \pi _{\phi } \cr
               e^{6\alpha }m^2 \phi & \pi _{\alpha }}\right)
  \left(\matrix{\pi _{y_n} - {y_n \over \Sigma} \cr
                \pi _{z_n} \cr}\right) \cr}  \eqno(3.23)
$$
where $\Sigma = -K_n \pi _{\alpha } + e^{6 \alpha }m^2 \phi \pi _{\phi }$. The
linear Hamiltonian constraint then implies that, imposed as a quantum
constraint,
$$
\pi _{y_n}\Psi (y_n,z_n) =0  \eqno(3.24)
$$
and so $\Psi $ is independent of
$y_n$. Therefore the true degree of freedom has been isolated - the  wave
function is found to depend only on the single physical variable
$$
z_{n}=\pi _{\phi }s_n + \pi _{\alpha }f_n = a^2 (\phi ' s_n -{\cal H}f_n )
\eqno (3.25)
$$
and on the
background variables $\tilde a$ and $\tilde \phi$ (in the rest of the paper we
will drop the tilde on $a$ and $\phi$). The expression for the Hamiltonian for
the modes $z_n$ is rather complicated and is shown only in Appendix A.

We can find the  the wave function for the scalar perturbations
in terms of the real degree of freedom by using the semiclassical approximation
to the path integral expression for the wave function as in the graviton case
$$
\psi^s (a,\phi, z_{n}) \sim C(a,\phi) \exp (- I^{cl}_{_E})
\eqno(3.26)
$$
The Euclidean action of the saddlepoint contribution to the path
integral is a boundary term (since the action is quadratic) given by
$$
I^{cl}_{n}=( M z_n z_n' - N z_n^2 )|^{\eta^f_E}_{\eta^i_E}
\eqno (3.27)
$$
for
$$\eqalign{
M &= {(n^2-4)\over 2 [ (n^2-4){a'}^2 + 3 a^2{\phi'}^2]}\cr
N &= {1\over 4 M U a^3}
     \Big[ K_n ( 2a^4 -3a^6m^2\phi^2 + 3{(n^2-1)\over (n^2-4)} a^4{\phi'}^2 )
      + a^{12}m^4\phi^2 + 3a^9\phi\phi'a'\Big]\cr
U &= K_n aa' + a^8m^2\phi\phi'             \cr}
$$
and the derivatives here are with respect to Euclidean conformal time.

It is difficult to find solutions of the equation for $z_n$. It is easier to
return to the original variables and pick a particular gauge.  In order to
study
the scalar perturbations  we shall choose the gauge $b_{n}=k_{n}=0$, which is
known as the longitudinal gauge. Once the result has been obtained in this
gauge
it will be easy to recast it in terms of the true degree of freedom $z_n$ and
therefore in a gauge invariant way. Alternatively, we could use the gauge
invariant variables of Bardeen.$^{\refbar}$ Their relationship with the
formalism used here is  described in appendix B.

In the $b_{n}=k_{n}=0$ gauge we have the equations of motion (in Lorentzian
time)
$$
a_{n}''+3{\cal H}a_{n}'+(3m^{2}\phi ^{2} a^{2} -2)a_{n}=3(m^{2}\phi
a^{2}f_{n}-\phi 'f_{n}')
\eqno(3.28)
$$
$$
f_{n}''+2{\cal H}f_{n}'+(n^{2}-1+m^{2}a^{2})f_{n}=2m^{2}\phi a^{2}a_{n}-4
\phi ' a_{n}'
\eqno(3.29)
$$
and the constraints
$$
a_{n}' +{\cal H}a_{n}=-3\phi 'f_{n}
\eqno(3.30)
$$
$$
a_{n}(n^{2}-4-3\phi '^{2})=3\phi ' f_{n}'+3m^{2}\phi a^{2}f_{n}+9{\cal
H}\phi 'f_{n}.
\eqno(3.31)
$$
Equations (3.28-30) are just (A.7), (A.11), (A.8), noting that $g_n=-a_n$
in this gauge. The last equation follows from (A.14), (3.30) and the
background constraint.  These equations are not independent, the  first one can
be obtained by taking a derivative of the first constraint and using the second
equation and the background equation of motion. Equations (3.28), (3.30) and
(3.31) can be combined to give the decoupled equation of motion for $a_{n}$:
$$
a_{n}'' +2({\cal H}-{\phi '' \over \phi '})a_{n}'+(2{\cal H}'-2{\cal H}
{\phi '' \over \phi '}+n^{2}+3)a_{n}=0
\eqno(3.32)
$$
This equation is useful in the inflationary era where $\phi '\neq 0$.  It is
also useful in the limit where the curvature of the 3-space can be neglected as
we can solve it explicitly in either the adiabatic or non-adiabatic regime (see
[\refmuk91]).   Once we have a solution for $a_n$, we can also find $f_n$
using the constraint equations (3.30) or (3.31), and therefore the real degree
of freedom $z_n$. In the region near the maximum expansion it is much harder to
solve (3.32) and we return to (3.28-3.31).

\vskip 1truecm

{\bf \sl (b) No Boundary Proposal Mode Function}

\vskip 0.5truecm

Let us now construct the solutions of (3.28-3.32) selected by the no boundary
proposal.  We  focus only on modes which go outside the Hubble radius during
inflation.   These are the ones which get excited  by the varying gravitational
field.  The very high frequency modes remain adiabatic throughout the history
of the universe, so their arrows of time will agree with the cosmological one.
As in the graviton case we divide the background saddle-point
4-geometry  into an approximately Euclidean section, followed by an
inflationary
one which finally turns into dust. We have however to take into account the
detailed behaviour of the background scalar field $\phi$ as it couples directly
to the perturbations.  We first find the regular
Euclidean solutions and match them up to the ones in the inflationary phase.
This can be done by analytic continuation.  In the inflationary era the modes
oscillate for a while until they leave the Hubble radius.  At that point we
match them to nonadiabatic solutions. Finally, the inflationary era comes to an
end when $\phi $ becomes small and starts oscillating, behaving like a dust
background.  At this point we match on the solutions for the dustlike
phase.  It turns out that for the Euclidean and inflationary solutions the
right hand terms in (3.29) are negligible.  We can solve for the scalar field
modes $f_n$ and calculate $a_n$ from an integral version of the constraint
(3.30) and check that this agrees with the approximate solutions of (3.32).  If
these terms were negligible during the whole of the dust era the modes would
oscilllate adiabatically around the maximum expansion as in the graviton case.
However we show that these terms do contribute to a monotonically increasing
amplitude of the scalar field perturbations  around maximum expansion.

The no boundary proposal requires that the matter fields in the path integral
be
regular, so in the semiclassical approximation we look for solutions to the
Euclidean perturbation equations which are regular as $\tau \rightarrow 0$.
The
regularity condition  requires that $f_n$ and $a_{n}$ vanish as $\tau
\rightarrow
0$. For $n \gg 1$, the dominant terms of equation (3.32) are the second
derivative of $a_n$ and  $-n^{2}$ times $a_{n}$ and one can construct a WKB
solution.  The approximate Euclidean solution selected by the no
boundary proposal is
$$
a_{n} \approx A{\phi ' \over a}e^{n\eta _{_E}},\qquad
f_n \approx  -{An\over 3} {e^{n\eta _{_E}}\over a}
\eqno(3.33)
$$
for some complex constant $A$.
Here, the conformal time $\eta _{E}=0$ corresponds to the juncture of Euclidean
and Lorentzian spacetimes. Continuing the regular Euclidean solution into the
Lorentzian section, taking $\eta_{_E}\rightarrow i\eta$, gives
$$
a_{n} \approx {1\over3}imAe^{in\eta }  \qquad
f_n \approx -{An\over 3} {e^{in\eta }\over a}
\eqno(3.34)
$$
where we have used $\phi '/a = im/3$ during inflation (dash now denotes
Lorentzian time derivative). The analytical  continuation holds into
the inflationary era as long as the wavelength is smaller than the  Hubble
radius, i.e. $n \gg {\cal H}$. By this time inflation has begun and we can
match
onto the inflationary solutions. When the modes move outside the Hubble radius
the modes $a_n$ and $f_n$ stop oscillating. They both have decaying and growing
modes (the latter would be constant in the limit of exact de Sitter space). As
the universe inflates only the slowly growing mode remains$^{\refmuk91 }$ so
that
$$
a_n \approx {D\over \phi^2 } \qquad f_n \approx {D\over \phi}
\eqno (3.35)
$$
where $D={1\over3}miAe^{in\eta _{H}}(\phi _{H}^{2} +{in\phi _{H} \over
ma_{H}})$
is a constant depending on the detailed matching of the modes when they cross
the Hubble radius at the time $\eta_H$.  This solution is valid until the
background scalar field  decrease to $\phi \sim 1$. Figure 3 depicts the
behaviour of $a_n$ during inflation and the beginning of the dust phase.

Eventually inflation ends and the background scalar field begins to oscillate.
We expect that the the background will behave effectively as a dust-filled
universe (see equation (2.16)) for perturbation modes with physical wavelengths
much larger than the scalar field compton wavelength ($n \ll ma$) since the
pressure of the oscillating scalar field averages to zero over that wavelength
scale. Therefore the metric perturbations will behave like those of a pure dust
universe (see, e.g. [\refmuk91]). This is indeed what is found below.

During inflation the Hubble radius $H^{-1}$ is roughly constant but as the
universe evolves in the dust era the Hubble radius starts growing.  When it
becomes larger than the compton wavelength $ 1/ m a$, the dominant term  in
(3.29) is $m^2a^2$.  The perturbation of the scalar field will start
oscillating
again.  In this early stage of the dust era when the curvature of the 3-surface
is negligible it can  be shown that the $f_n$ oscillate exactly in phase with
$\phi'$ as follows:
$$
f_n \approx -{\phi ' \over a}\int d\eta aa_n.
\eqno(3.36)
$$
This will remain true in later stages of the dust era as long as $n <ma_e$.
This
condition ensures that the phase of $f_n$ obtained by integrating (3.29) does
not
differ appreciably from that of $\phi '$.  Using (3.36) together with (3.30) we
can establish that the metric perturbation $a_n$, time averaged over one
oscillation period of $\pi /m$, is growing. The small oscillations around this
average arise because the background energy momentum tensor is not exactly that
of dust but that of an oscillating scalar field. The averaged gravitational
perturbation $a^A_n$ can be calculated by taking the derivative of the averaged
version of (3.30) to obtain the differential equation
$$
  {a^A_n}'' +3{\cal H}{a^A_n} ' -2a^A_n = 0.   \eqno(3.37)
$$
The general solution is a linear combination of the solutions
$$  a_{n}
^{\rm anti}\approx {{\sin \eta _{d}} \over {(1-\cos\eta _{d})^{3}}}
\eqno (3.38)
$$
and
$$
a_{n} ^{\rm sym}\approx  {{2\sin ^{2} \eta _{d}-6(\eta _{d}-\pi )\sin \eta
_{d}-8\cos\eta _{d} +8}
         \over {(1-\cos \eta _{d})^{3}}} .
\eqno(3.39)
$$
The conformal time is defined with the new origin at the beginning of the dust
phase ($\eta _{d} \approx 0$). These solutions are antisymmetric and
symmetric with respect to the maximum of expansion ($\eta _d = \pi $) and are
the
same solutions found for perturbations in a pressureless perfect fluid
universe,
as expected. Both solutions diverge like $ \eta _{d}^{-5}$ in the beginning of
the dust era  as $\eta_{d}\rightarrow 0$. There is however a regular solution,
given by $a_n^{\rm reg}:= a_n^{\rm symm}-6\pi a_n^{\rm anti}$, which approaches
a
constant in this limit.  At the end of inflation the $a_n$ picked out by the no
boundary proposal are small as seen from (3.35).  Therefore the regular
solution
is the one selected by the no boundary proposal and this is asymmetric in the
dust era: the perturbation amplitude steadily increases with time. Matching the
solutions for the dust era to (3.35) shows that during the dust era
$$
  a_n \approx D a_{n} ^{\rm reg} \eqno (3.40)
$$
We can now use (3.36) to see that $f_n$ is oscillating with monotonically
increasing amplitude throughout the dust era:
$$
f_n\approx -{D\phi ' \over (1-\cos \eta _d)^2}[4\eta _d - 6\sin \eta _d +2\eta
_d \cos \eta _d] \eqno (3.41)
$$

With these solutions we can construct the
wave function  (3.26).  When the background saddle-point is approximatively
Lorentzian, the no-boundary  wave function for the scalar perturbation is
$$
\psi ^s (z_n) \sim C(a,\phi ) \
\exp -i \Big( M({\mu _n' \over \mu _n})
+ N(a,\phi) \Big)z_n^2 .
\eqno(3.42)
$$
where $M$ is given in (3.27) and $\mu _n$ is the modefunction for $z_n$.  It is
a solution of the equation of motion for $z_n$ picked out by the no-boundary
proposal.  It is explicitly given by the function $a_n$ and $f_n$ using (3.25)
with $z_n$ replaced by   $\mu_n$.  From the solution of $a_n$ and $f_n$ we can
see that it is clearly asymmetric about the time of maximum expansion.
Considering points placed symmetrically about the maximum of expansion, the
background will be the same at both points so that the asymmetry in the
modefunction manifests itself as an asymmetry in the wavefunction. The variance
of $z_n$ is proportional to the modulus of $\mu _n$   and is therefore
asymmetric with respect to the time of maximum expansion. We therefore conclude
that the wavefunction predicts the continuing growth of low frequency scalar
perturbations even when the universe begins to recollapse.

This result alone provides a time asymmetry so long as the modes stay in a
regime where they can be treated in a linear approximation. However, most modes
will also enter a nonlinear regime well before the maximum expansion occurs.
When this occurs the interaction terms in the Hamiltonian will become important
and hence the coarse-grained entropy will increase throughout the evolution.

Considering the stress tensor in the gauge-invariant formalism (see e.g.
[\refmuk91 ]), we can show that the density contrast is
$$
{\delta \rho \over \rho} \approx {2a \over 3a_m}[(n^2-4)a_n -9{\cal H}\phi '
f_n]  \eqno(3.43)
$$
Modes will cross the horizon (${\cal H} \sim n$) when $\eta_d \sim 1/n$, and
the recent COBE results$^{\refSmoot }$ indicate that the density contrast
at this time is of order $10^{-5}$. Using equations (3.40) and (3.41) in
(3.43), we find that the constant $D$ is of order $10^{-5}$. At later
times, only the first term in (3.43) is important and we find that the
density contrast behaves like
$$
{\delta \rho \over \rho} \approx 10^{-7} n^2 \eta_d^2  \eqno(3.44)
$$
Consequently, when the density contrast is of order unity we expect
nonlinearity to be the dominant feature and this occurs for $\eta_d^2 \ge
10^7/n^2$. Modes with $n \ge 1000$ will therefore enter a nonlinear phase
before
they reach the maximum and the coarse-grained entropy for these modes will
grow.

\proclaim 4) Conclusion.

In this paper we have investigated the consequences of the no-boundary proposal
for the arrow of time.  In particular  we have investigated the behaviour of
small metric and matter perturbations around a homogeneous isotropic
background.  The no-boundary proposal predicts classical evolution with an
inflationary era followed by a dustlike era.  We found that perturbation modes
are in their ground state at the beginning of the inflationary era. This can be
interpreted as a statement that the universe is born in a low entropy state.
Modes which leave the Hubble radius during inflation become excited then
subsequently evolve in various ways in the dustlike era.

We find that gravitons oscillate adiabatically
for most of the dustlike era and consequently the amplitude of their
oscillations is time symmetric with respect to the point of maximum scale
factor. However, looking at the physical scalar degrees of freedom we find that
those which have been excited by superadiabatic amplification during
inflation have a time asymmetric evolution with respect to the maximum. In
particular, the variance of the scalar modes predicted by the wavefunction is
different at the same value of the scale factor before and after the maximum.

Thus we find that the wavefunction of the universe distinguishes between
symmetrically placed points on either side of maximum volume.  The expanding
phase has a smaller amplitude of the variance in the low frequency scalar modes
than does the corresponding point during the collapsing phase. In other words,
the thermodynamic arrow coincides with the cosmological arrow before the
maximum,
but points in the opposite direction after the maximum. This is true for all
the
lowest frequency modes, so that they induce a well-defined thermodynamic arrow
of time. Amongst the modes which display this nonadiabatic behaviour,
higher frequency modes will enter a nonlinear regime during the expansion and
consequently produce a growing coarse-grained entropy throughout
expansion and recontraction, and hence also create a thermodynamic arrow of
time.

\proclaim Acknowledgments.

We would like to thank Nathalie Deruelle and Carsten Gundlach for useful
conversations.
R.L. acknowledges Peterhouse,Cambridge, for financial support. G.W.L.
acknowledges the Science and Engineering Research Council for
supporting this work.

\magnification=\magstep1
\openup 1\jot
\proclaim Appendix A) Action and field equations.

In this appendix we reproduce the action and field equations of the perturbed
FRW model driven by a massive minimally coupled scalar field from
ref.[\refhaha85].
The homogeneous part of the Einstein-Hilbert Lagrangian is
$$\displaylines{
L_0=-{1\over2} N_0a^{3}[{{\dot a}^{2} \over N^{2}a^{2}}
-{1 \over a^{2}}-{{\dot \phi }^{2} \over N^{2}}
+m^{2}\phi ^{2}]\hfill (A.1)\cr}
$$
The second order perturbation of the Einstein Hilbert Lagrangian is
$$\displaylines{
L_g^n=\half aN_0 \Bigg\{
{1\over 3}(n^2-{5\over 2})a_n^2 +{(n^2-7)\over 3}{n^2-4)\over(n^2-1)}b_n^2
- 2(n^2-4)c_n^2 -(n^2+1)d_n^2 +{2\over 3}(n^2-4)a_nb_n \hfill\cr}
$$
$$\ \ \ \ \ \ \
+{2\over 3}g_n[(n^2-4)b_n + (n^2+\half)a_n]
+{1\over N_0^2} \Big[-{1\over 3(n^2-1)}k_n^2 + (n^2-4)j_n^2\Big]\Bigg\}
$$
$$\displaylines{ \ \ \ \ \ \ \
+\half{a^3\over N_0}
\Bigg\{ -\dot a_n^2 + {(n^2-4)\over(n^2-1)}\dot b_n^2
       + (n^2-4)\dot c_n^2 + \dot d_n^2
+g_n[2{\dot a\over a}\dot a_n + {\dot a^2\over a^2}(3a_n-g_n)]
\hfill\cr}
$$
$$
+{\dot a\over a}\Big[-2a_n\dot a_n +8{(n^2-4)\over(n^2-1)}b_n\dot b_n
                     +8(n^2-4)c_n\dot c_n + 8d_n\dot d_n \Big]
$$
$$
+{\dot a^2\over a^2}\Big[-{3\over 2}a_n^2 + 6{(n^2-4)\over(n^2-1)}b_n^2
                         +6(n^2-4)c_n^2 + 6d_n^2 \Big]\ \ \ \ \ \ \ \ \ \ \
$$
$$
+{1\over a}\Big[{2\over 3}k_n\Big\{-\dot a_n - {(n^2-4)\over(n^2-1)}\dot b_n
                                    +{\dot a\over a} g_n\Big\}
                 -2(n^2-4)\dot c_n j_n \Big]
\Bigg\}
\eqno (A.2)
$$
The perturbation of the matter Lagrangian gives:
$$\displaylines{
L_m^n=\half N_0a^3\Bigg\{
{1\over N_0^2}(\dot f_n^2+6a_n\dot f_n\dot\phi) -m^2(f_n^2 +6a_nf_n\phi)
-{1\over a^2}(n^2-1)f_n^2 +{\dot\phi^2\over N_0^2}g_n^2\hfill\cr}
$$
$$
+{3\over 2}\Big[{\dot\phi^2\over N_0^2}-m^2\phi^2\Big]
   \Big[ a_n^2-4{(n^2-4)\over(n^2-1)}b_n^2 -4(n^2-4)c_n^2 -4d_n^2\Big]
$$
$$
-g_n\Big[2m^2f_n\phi + 3m^2a_n\phi^2 + 2{\dot f_n\dot\phi\over N_0^2}
+3{a_n\dot\phi^2\over N_0^2}\Big] -2{1\over aN_0^2}k_nf_n\dot\phi\Bigg\}.
\eqno (A.3)
$$

The field equations necessary to calculate the saddle point approximation are
given below.
{}From (A.1) we find the equations obeyed by the homogeneous background
fields. The homogeneous scalar field $\varphi_0$  obeys
$$\displaylines{
N_0 {d\over dt} \Big [{1\over N_0}{d\varphi_0\over dt} \Big ]
+3 {d a\over a dt}{d\varphi_0\over dt}  + N_0 m^2 \varphi_0 =
{\rm quadratic \ terms,}
\hfill (A.4)\cr}
$$
and the scale factor $a$ obeys
$$ \displaylines{
N_0 {d\over dt} \Big [{1\over N_0 a}{da\over dt}\Big ]
+ 3 \dot\varphi^2_0 -{N^2_0\over a^2}
- {3\over 2}(-{\dot a^2\over a^2} +\dot\varphi^2_0 - {N^2_0\over a^2}
             + N^2_0 m^2 \phi^2)
=  {\rm quadratic \ terms,}
\hfill (A.5)\cr}
$$
The background variables $a$, $\varphi_0$ and their momenta are subject to the
 constraint
$$
\displaylines{
-{\dot a^2\over a^2N^2_0} + {\dot\varphi_0\over N^2_0} - {1\over a^2} +
m^2\varphi^2_0 = {\rm quadratic \ terms.}
\hfill (A.6)\cr}
$$

Let us now turn to the equation of motion of the small inhomogeneities.
Variations with respect to $a_n,b_n,c_n,d_n$ and $f_n$ give the following
second order field equations:
$$\displaylines{
N_0 {d\over dt} \Big [{a^3\over N_0 }{da_n\over dt}\Big ]
+{1\over 3}(n^2-4) N^2_0 a(a_n+b_n)
+ 3 a^3(\dot\varphi_0\dot f_n - N^2_0 m^2\varphi_0 f_n)\hfill\cr}
$$
$$
= N^2_0 [ 3a^3m^2\varphi^2_0 - {1\over 3}(n^2+2)a]g_n
+ a^2 \dot a \dot g_n
-{1\over 3}N_0 {d\over dt} \Big [{a^2 k_n\over N_0 }\Big ],
\eqno (A.7)
$$
$$\displaylines{
N_0 {d\over dt} \Big [{a^3\over N_0 }{db_n\over dt}\Big ]
-{1\over 3}(n^2-1) N^2_0 a(a_n+b_n)=
{1\over 3}N^2_0 (n^2-1)a g_n
+{1\over 3}N_0 {d\over dt} \Big [{a^2 k_n\over N_0 }\Big ],
\hfill (A.8)\cr}
$$
$$\displaylines{
N_0 {d\over dt} \Big [{a^3\over N_0 }{dc_n\over dt}\Big ]=
{d\over dt} \Big [{a^2 j_n\over N_0 }\Big ],
\hfill (A.9)\cr}
$$
$$\displaylines{
N_0 {d\over dt} \Big [{a^3\over N_0 }{d\, d_n\over dt}\Big ]
+(n^2-1)N^2_0a d_n =0,
\hfill (A.10)\cr}
$$
and
$$\displaylines{
N_0 {d\over dt} \Big [{a^3\over N_0 }{df_n\over dt}\Big ]
+ 3a^3\dot\varphi_0 \dot a_n + N^2_0[m^2a^3 + (n^2-1)a]f_n \hfill\cr}
$$
$$
= a^3(-2N^2_0m^2\varphi_0 g_n + \dot\varphi_0\dot g_n -{\varphi_0 k_n\over a}).
\eqno (A.11)
$$
Variations with respect to $k_n,j_n$ and $g_n$ lead to the constraints
$$\displaylines{
\dot a_n + {(n^2-4)\over(n^2-1)} \dot b_n + 3f_n \dot\varphi_0 =
{\dot a g_n\over a} - {k_n \over a (n^2-1)},
\hfill (A.12)\cr}
$$
$$\displaylines{
\dot c_n = {j_n\over a}
\hfill (A.13)\cr}
$$
and
$$\displaylines{
3a_n ( \dot\varphi^2_0 -{\dot a^2\over a^2})
+ 2 (\dot\varphi_0\dot f_n- {\dot a\dot a_n\over a})
+ N^2_0 m^2 (2f_n\varphi_0 + 3 a_n \varphi^2_0)
-{2N^2_0\over 3a^2}  [(n^2-4)b_n + (n^2 + {1\over 2} ) a_n]\hfill\cr}
$$
$$
= {2\dot a k_n\over 3a^2} + 2 g_n ( \dot\varphi^2_0 -{\dot a^2\over a^2}).
\eqno (A.14)
$$

We also give the perturbation Hamiltonian in terms of the real degrees of
freedom:

$$
H^n_2(z_n,\pi_{z_n})= A\pi^2_{z_n} + B z_n\pi_{z_n} + C z^2_n
$$
with
$$\eqalign{
A &= {1\over 2}\big( a{\dot a}^2+{3a^3\dot\phi^2 \over (n^2-4)} \big) \cr
B &= -{1\over2U}
     \Big[ K(2a - 3a^3m^2\phi^2 - 3{(n^2-1)\over(n^2-4)}a^3{\dot\phi}^2)
           + a^9m^4\phi^2 - 3a^8 m^2\phi\dot\phi\dot a  \Big]\cr
C &= -{1\over2U^2}
     \Big[ -{3(n^2-1)K^3\over (n^2-4) a^3} + a^3m^2K^2 -5a^9m^4\phi^2K
           +12a^{15}m^4\phi^2{\dot\phi}^2
     \Big]\cr
K &= -3a^6{\dot\phi}^2 -{(n^2-4)\over 3}a^4;  \qquad\qquad
U = -Ka^2\dot a - a^9m^2\phi\dot\phi        \cr}
\eqno (A.15)
$$

\def\ch{{\cal H}}
\openup 1\jot
\proclaim Appendix B: Relation to the gauge invariant formalism.

There has recently been much interest in the gauge invariant
formalism$^{\refbar,\refmuk91}$ which cast the variables of the theory (the
scalar perturbations of the gravitational and scalar fields) into ones which
are
invariant under infinitesimal gauge transformations.  In this appendix we
relate
the different harmonics in equations (3.4) and (3.5) to the gauge invariant
variables (in particular we shall follow the approach of [\refmuk91]).

Mukhanov et al. define the time and space dependent scalar metric perturbations
as
$$
ds^2 = a^2(\eta) \Big\{ (1+2\phi)d\eta^2 - 2B_{|i} dx^i d\eta +
[(1-2\psi)\gamma_{ij} + 2E_{|ij}] dx^idx^j \Big \}
\eqno (B.1)
$$
and the scalar field perturbations
$$
\varphi(\vec x,t) = \varphi_o(t) + \delta\varphi(\vec x, t)
\eqno(B.2)
$$

The above variables are related to the modes perturbations used in
this paper in the following way
$$
\eqalign{
\phi &= \sum_n {g_n Q^n \over \sqrt{6}} \cr
\psi &= \sum_n {-(a_n+b_n) Q^n \over \sqrt{6}} \cr
B &= \sum_n {k_n Q^n \over (n^2-1)\sqrt{6}} \cr
E &= \sum_n {3 b_n Q^n \over (n^2-1)\sqrt{6}} \cr}
\eqno (B.3)
$$
where we have suppressed in the sum the indices ${lm}$ corresponding to the
angular momentum.

Under a general linear gauge transformation of the form
$$
\eqalign{
\eta \rightarrow &\tilde\eta = \eta + \xi^0(\eta,\vec x)\cr
x^i\rightarrow & \tilde x^i = x^i + \gamma^{ij}\xi_{|j}(\eta,\vec x)\cr}
\eqno (B.4)
$$
the scalar perturbations transform as
$$
\eqalign{
\tilde\phi & = \phi - {a^\prime\over a}\xi^0 - \xi^{0^\prime} \cr
\tilde\psi & = \psi + {a^\prime\over a}\xi^0 \cr
\tilde B &= B + \xi^0 - \xi^\prime \cr
\tilde E &= E - \xi \cr
\tilde{\delta\phi} &= \delta\phi -\varphi_o^\prime \xi^{0^\prime}.\cr}
\eqno (B.5)
$$
The idea of the gauge invariant formalism is to make a linear combination
of the different scalar perturbations such that the resulting variables are
independent of the gauge.  A possible choice is
$$
\eqalign{
\Phi &= \phi + {1\over a}[(B-E^\prime)a]^\prime \cr
\Psi &= \psi -  {a^\prime\over a}(B-E^\prime)  \cr
\delta\varphi^{(gi)} &= \delta\varphi + \varphi^\prime_0(B-E^\prime) \cr}
\eqno (B.6)
$$

These gauge invariant quantities obey the following equations:
$$
\eqalignno{
\nabla^2\Phi - 3 \ch\Phi^\prime - (\ch^\prime +2\ch^2 - 4 K)\Phi &=
 {3\ell^2\over 2} (\varphi^\prime\delta\varphi^{(gi)\prime}
+ V_{,\varphi}a^2 \delta^{(gi)}), & (B.7)\cr
\Phi^\prime + \ch \Phi &=
{3\ell^2\over 2}\varphi^\prime\delta\varphi^{(gi)}, & (B.8)\cr
\Phi^{\prime\prime} + 3 \ch\Phi^\prime + (\ch^\prime +2\ch^2)\Phi &=
 {3\ell^2\over 2} (\varphi^\prime\delta\varphi^{(gi)\prime}
- V_{,\varphi}a^2 \delta\varphi^{(gi)}), & (B.9)\cr}
$$
which are the gauge invariant versions of the $\delta G^0_0=8\pi G\delta
T^0_0$,
$\delta G^0_i=8\pi G\delta T^0_i$ and $\delta G^i_j=8\pi G\delta T^i_j$
equations  and
$$
\delta\varphi^{(gi)\prime\prime} +2\ch\delta\varphi^{(gi)\prime}
-\nabla^2\delta\varphi^{(gi)} + V_{,\varphi\varphi}a^2 \delta\varphi^{(gi)}
-4\varphi^\prime_o\Phi^\prime + 2V_{,\varphi}a^2\Phi = 0.
\eqno (B.10)
$$
is the gauge invariant version of the scalar field equation.

In the longitudinal gauge ($B=k_n=0$ and $E=b_n=0$) used in this paper the
gauge
variables reduce to $\Phi=\phi$, $\Psi=\psi$ and $\delta\varphi^{(gi)} =
\delta\varphi$ and if we expand them in harmonics on the 3-sphere
$\Phi_n=g_n/\sqrt{6}$, $\Psi_n=-a_n/\sqrt{6}$ and  $\delta\varphi^{(gi)} =
f_n/\sqrt{6}$.  Indeed it is easy to see  that equations (B.9) and (B.10) are
equivalent to (3.31) and (3.32) respectively, and that the  constraint (B.8)
is equivalent to (3.33).

\openup 1\jot

\proclaim References.

\item{ [\refcpt]} R. F. Streater and A. S. Wightman, PCT, Spin Statistics
and All That, Benjamin, New York (1964).

\item{ [\refKmes]} J. H. Christenson, J. W. Cronin, V. L. Fitch \& R.
Turley, Phys. Rev. Lett. {\bf 13}, 138 (1964).

\item{ [\refqgcpt]} T. Banks, Nuc. Phys. {\bf B249}, 332 (1985).

\item{ [\refHH83]}  J. B. Hartle \& S. W. Hawking, Phys. Rev. {\bf D28} 2960
(1983).

\item{ [\refSmoot]} G. Smoot et al., Astrophys. J. {\bf 396}, L1 (1992).

\item{ [\refH85]} S.W. Hawking,  Phys. Rev. {\bf D32}, 2489 (1985).

\item{ [\refHW85]} S.W. Hawking \& Z.C. Wu Phys. Lett. {\bf B107}, 15 (1985).

\item{ [\refPa85]} D. Page,  Phys. Rev. {\bf D32}, 2496 (1985).

\item{ [\refLa88]} R. Laflamme, Time and Quantum Cosmology,  Ph.D. thesis,
Cambridge University (1988).

\item{[\refLS87]}  R. Laflamme \&  E.P.S. Shellard,  Phys. Rev. {\bf D35}, 2315
(1987).

\item{[\refHalHar]} J.J. Halliwell \& J.B. Hartle Phys. Rev. {\bf D41},
1815 (1990)

\item{[\refLy92]} G. Lyons, Phys. Rev. {\bf D46}, 1546 (1992).

\item{[\refHaL90]} S. Habib \& R. Laflamme, Phys.Rev. {\bf D42}, 4056
(1990).

\item{[\refli63]}  E.M. Lifschitz \& I.M. Khalatnikov, Adv.Phys. {\bf 12},
 185 (1963).

\item{[\refge78]}  U.H. Gerlach \&  U.K. Sengupta, Phys.Rev. {\bf D18},
1773 (1978).

\item{[\refhaha85]}  J.J. Halliwell \&  S.W. Hawking, Phys.Rev. {\bf D31}, 1777
(1985).

\item{ [\refG74]}  L. P. Grishchuk, Zh. Eksp. Teor. Fiz. {\bf 67}, 825 (1974).

\item{ [\refLaL91]} R. Laflamme \& J. Louko, Phys. Rev. {\bf D43}, 2730
(1991)

\item{ [\refgr89]}  L. P. Grishchuk and Yu.V. Sidorov;
Class. \& Quant.Grav. 6, L161  (1989).

\item{ [\refwa88]} I. Shirai \&  S. Wada, Nuc.Phys. {\bf B303}, 728 (1988).

\item{ [\refbar]} J. Bardeen, Physical Review {\bf D 22}, 1882 (1980)

\item{ [\refmuk91]}  V.F. Mukhanov, H.A.  Feldman \&  R.H. Brandenberger,
Theory of cosmological perturbations, Brown University report,
Brown-HET-796 (1991).

\bye